\documentclass[conference]{IEEEtran}

\usepackage{dsfont}
\usepackage{gensymb}
\usepackage{amsmath}
\usepackage{amsfonts}
\usepackage{epsfig}
\usepackage{amssymb}
\usepackage{cite}
\usepackage{hhline}
\usepackage{multirow}
\usepackage{framed}
\usepackage{xcolor}
\usepackage{lipsum}
\usepackage{bm}

\usepackage{makecell}
\usepackage{color}
\usepackage{graphicx}
\usepackage{subfigure}

\usepackage{tikz}
\usepackage{pgfplots}

\makeatletter
\newcommand{\vast}{\bBigg@{3}}
\newcommand{\Vast}{\bBigg@{4}}
\makeatother
\renewcommand{\IEEEQED}{\IEEEQEDopen}
\newtheorem{defin}{Definition}%[Section]

\newtheorem{result}{Result}

\begin{document}

\title{On the Accuracy of Interference Models in \\ Wireless Communications}
\author{\IEEEauthorblockN{Hossein Shokri-Ghadikolaei\IEEEauthorrefmark{2}, Carlo Fischione\IEEEauthorrefmark{2}, and Eytan Modiano\IEEEauthorrefmark{3}}
\IEEEauthorblockA{\IEEEauthorrefmark{2}Electrical Engineering, KTH Royal Institute of Technology, Stockholm, Sweden \\\IEEEauthorrefmark{3}Laboratory for Information and Decision Systems, Massachusetts Institute of Technology, Cambridge, USA\\
{Emails: hshokri@kth.se, carlofi@kth.se, and modiano@mit.edu}}}

\maketitle

\begin{abstract}
We develop a new framework for measuring and comparing the accuracy of any wireless interference models used in the analysis and design of wireless networks. Our approach is based on a new index that assesses the ability of the interference model to correctly predict harmful interference events, i.e., link outages.  We use this new index to quantify the accuracy of various interference models used in the literature,  under various scenarios such as Rayleigh fading wireless channels, directional antennas, and blockage (impenetrable obstacles) in the network.  Our analysis reveals that in highly directional antenna settings with obstructions, even simple interference models (e.g., the classical protocol model) are  accurate, while with omnidirectional antennas, more sophisticated and complex interference models (e.g., the classical physical model) are necessary. Our new approach makes it possible to adopt the appropriate interference model of adequate accuracy and simplicity in different settings.
\end{abstract}

\begin{IEEEkeywords}
Interference model, performance evaluation, protocol design, millimeter wave communications.
\end{IEEEkeywords}

\section{Introduction}

Due to the shared nature of a wireless media, interference plays a critical role in  the performance of wireless networks, where the intended signal is combined with other undesired signals transmitted on the same (time, frequency, spatial) channel. The receiver typically decodes the received signal by treating the interference as noise, though advanced receivers may be able to cancel some parts of the interference. Due to the randomness of the channel attenuation and the interferers, successful decoding at the receiver is a random event whose probability depends on the desired signal strength, the ambient noise level accumulated over the operating bandwidth, and the interfering signals strength. Signal-to-interference-plus-noise ratio (SINR) is a common metric to evaluate outage probability (or probability of successful decoding) of a transmission. However, evaluating the outage probability using the SINR model is complex as it depends on the transmission powers, unknown random channel attenuation, medium access control (MAC) protocol used, and the network topology, which is often unknown. Thus, although the SINR interference model is very accurate, using it for the design and analysis of wireless networks is challenging and often results in little insight.

There have been many attempts in the literature to design interference models that accurately capture the effect of interference, yet are tractable for the mathematical analysis. Among the most prominent models, introduced in the literature, are the protocol model of interference~\cite{gupta2000capacity}, the interference ball model~\cite{le2010longest}, and the physical model~\cite{gupta2000capacity}.

The \emph{protocol model} (PRM) is the simplest model, formalized by the seminal work of Gupta and Kumar~\cite{gupta2000capacity}. Under the PRM, an outage event occurs if the closest interferer is no farther than a certain distance from the receiver, called the interference range. The interference range depends on the received power from the intended transmitter and a minimum SINR threshold that allows successful signal decoding. Although the PRM is simple, especially for the protocol design and for the MAC layer performance analysis, it fails to capture the effect of interference aggregation (i.e., the sum of the interference power from multiple interferers). Thus, the PRM is generally considered to be overly simplistic.  Nonetheless, due to its mathematical tractability, the PRM has been extensively adopted for the analysis of MAC protocols and network  performance; e.g., transport capacity~\cite{gupta2000capacity}, delay~\cite{el2006optimal}, and collision probability~\cite{Shokri2015Transitional}.

The \emph{interference ball model} (IBM) attempts to alleviate the aforementioned limitation of the PRM by  considering the aggregated impacts of  near-field interferers, located no farther than a certain distance from the receiver. This model has been extensively adopted in performance evaluation and protocol design for wireless networks~\cite{weber2007transmission,le2010longest,di2014stochastic}. The IBM is more accurate than the PRM, but also more complex.

The most accurate and  complex interference model is the \emph{physical model} (PhyM),\footnote{Although PhyM may be mathematically more tractable than both PRM and IBM~\cite{Haenggi2013Stochastic} under very special network settings (e.g., homogenous Poisson field of interferers exhibiting Rayleigh fading channel), PRM and IBM are more favorable interference models for protocol design and for network optimization~\cite{le2010longest}.} formalized in~\cite{gupta2000capacity}, which considers the aggregated interference of all transmitters in the entire network. This interference model, also known as the SINR model, is adopted mostly at the physical layer for power control, capacity evaluation, and coverage analysis~\cite{gupta2000capacity,Haenggi2013Stochastic}.

Clearly there is a tradeoff between accuracy and complexity of interference models.  The proper choice of interference model depends on many parameters  such as the receiver design, antenna directionality, network topology, and the choice of medium access protocol.  To the best of our knowledge, there has been no systematic method \emph{for assessing the accuracy of various interference models, and for choosing the proper interference model for a given network scenario}. Most prior works have evaluated the accuracy of different interference models  qualitatively, without fully understanding the mutual impact of different parameters of the physical, medium access, and network layers. This qualitative analysis, however, is often overly simplistic, and may result in the use of interference models that are only marginally more accurate, yet significantly more complex than needed.  As we will show throughout this paper, in certain settings, even the simplest interference models are sufficiently accurate and can be used to provide significant insights on the network performance and design.

In this paper, we propose a systematic approach to rigorously quantify the accuracy of interference models in predicting the outage probability. We introduce an accuracy index that takes on real values between 0 and 1, where higher values correspond to higher accuracy. We evaluate this index for PRM and IBM under two example scenarios: (1) Rayleigh fading channel and omnidirectional communications; and (2) deterministic wireless channel, directional communications, and existence of impenetrable obstacles in the environment. The first scenario corresponds to conventional wireless networks~\cite{Haenggi2013Stochastic}, whereas the second scenario corresponds to emerging wireless technologies such as millimeter wave (mmWave) networks with highly directional antennas~\cite{Rangan2014Millimeter,shokri2015mmWavecellular}. Although the applications of the proposed index is general and goes much beyond the examples provided in this paper, we use these examples to investigate fundamental properties of this index and also the impact of various network parameters on the accuracy of IBM and PRM. In the first scenario, we derive a tractable closed-form expression for the accuracy index. We show that the accuracy of IBM monotonically increases with the interference range, at the expense of increased complexity.  In contrast, we show that  there is no such monotonic improvement in the accuracy of PRM. In the second scenario we show that both the PRM and IBM are significantly more accurate with directional antennas and channel blockage. Thus, the PRM can be used in the analysis of  mmWave networks.  This observation is very promising because the use of the PRM in mmWave networks can significantly improve mathematical tractability, with negligible loss in  accuracy of the performance analysis.

The rest of this paper is organized as follows. In Section~\ref{sec: system-model}, we present general assumptions, followed by the introduction of the new interference model accuracy index. We demonstrate the use of this index in Sections~\ref{sec: Rayleigh-noBlockage} and~\ref{sec: Rayleigh-Blockage}. Numerical results are  presented in Section~\ref{sec: numerical-results}, and concluding remarks are provided in Section~\ref{sec: conclusion}.

\section{Interference Model Accuracy Index}\label{sec: system-model}
We define a link as a transmitter and its intended receiver, where transmitter (receiver) $i$ refers to the transmitter (receiver) of link $i$. Denote by $\mathcal{I}_i$ the set of interferers of receiver $i$ (all transmitters excluding the intended transmitter\footnote{We assume that there is no interference cancellation, so all unintended transmitters act as potential interferers to any receiver. However, the framework proposed in this paper can be easily extended to cover the interference cancellation capability using similar technique adopted in~\cite{weber2007transmission}.}), by $p_i$ the transmission power of transmitter $i$, by $\sigma$ the power of white Gaussian noise, by $d_{ij}$ the distance between transmitter $i$ and receiver $j$, and by $g_{ij}^{c}$ the channel gain between transmitter $i$ and receiver $j$. The channel gain may include average attenuation at a reference distance, distance-dependent component, and fading components. We denote by $g_{ij}^{t}$ the antenna gain at transmitter $i$ toward receiver $j$, and by $g_{ij}^{r}$ the antenna gain at receiver $j$ toward transmitter $i$. Thus, the power received by receiver $j$ from transmitter $i$ is $p_i g_{ij}^{t} g_{ij}^{c} g_{ij}^{r}$, and the SINR at receiver $i$ is given by
\begin{equation*}
{\rm{SINR}}_{i} = \frac{p_i g_{ii}^{t} g_{ii}^{c} g_{ii}^{r}}{\sum\limits_{k \in \mathcal{I}_i} p_k g_{ki}^{t} g_{ki}^{c} g_{ki}^{r} + \sigma} \:.
\end{equation*}
Note that the SINR depends on the transmission powers, antenna patterns, and network topology. Let $\beta>0$ denote the SINR threshold corresponding to a certain target bit error rate. An outage on link $i$ occurs when ${\rm{SINR}}_{i}$ is lower then $\beta>0$.  Different interference models attempt to approximate the probability of outage by ignoring certain components of the interference.  In particular, an outage occurs under
\begin{itemize}
  \item PRM: if there is an interferer no farther than an interference range $r_{\text{PRM}} = (1+\Delta)d_{ii}$ of receiver $i$, where $\Delta$ is a constant real positive value;
  \item IBM: if its SINR due to all interferers located no farther than an interference range $r_{\text{IBM}}$ is less than $\beta$; and
  \item PhyM: if its SINR due to all interferers is less than $\beta$.
\end{itemize}

In order to present a unified view, we associate two random variables $a_{ij}^{\text{IBM}}$ and $a_{ij}^{\text{PRM}}$ to the link between each transmitter~$i$ and receiver $j \neq i$. $a_{ij}^{\text{IBM}}$ is set to $1$ if $d_{ij} \leq r_{\text{IBM}}$, and otherwise $0$. Similarly, $a_{ij}^{\text{PRM}}$ is set to ${+\infty}$ if $d_{ij} \leq (1+\Delta)d_{ii}$, and otherwise $0$. We can define a virtual channel gain as follows:
\begin{equation}\label{eq: general-model}
\begin{cases}
  g_{ij}^{\text{PRM}} = a_{ij}^{\text{PRM}} g_{ij}^{c}, & \mbox{for protocol model} \\
  g_{ij}^{\text{IBM}} = a_{ij}^{\text{IBM}} g_{ij}^{c}, & \mbox{for interference ball model} \\
  g_{ij}^{\text{PhyM}} = g_{ij}^{c}, & \mbox{for physical model} \:.
\end{cases}
\end{equation}
Now, the SINR at receiver $i$ under interference model $\mathrm{x}$ is given by
\begin{equation}\label{eq: protocol-model}
\gamma_{i}^{\mathrm{x}} = \frac{p_i g_{ii}^{t} g_{ii}^{c} g_{ii}^{r}}{\sum\limits_{k \in \mathcal{I}_i} p_k g_{ki}^{t} g_{ki}^{\mathrm{x}} g_{ki}^{r} + \sigma}  \:,
\end{equation}
where $\mathrm{x}$ is a label denoting $\text{PhyM}$, $\text{IBM}$, or $\text{PRM}$. Finally, there is an outage at receiver $i$ under model $\mathrm{x}$ if $\gamma_{i}^{\mathrm{x}} < \beta$.

Consider the physical model as the reference interference model. We define a binary hypothesis test, where hypotheses $H_0$ and $H_1$ denote the absence and presence of outage under the reference physical model PhyM. That is, for each receiver $i$, we have
\begin{equation}\label{eq: binary-hypothesis-test}
\begin{cases}
  H_0, & \mbox{if}~ \gamma_{i}^{\text{PhyM}} \geq \beta\\
  H_1, & \mbox{if}~ \gamma_{i}^{\text{PhyM}} < \beta \:.
\end{cases}
\end{equation}

We can consider a given interference model $\mathrm{x}$ as a detector of the outage events  at any given SINR threshold and network parameters. To evaluate the performance of this detector compared to the reference model $\text{PhyM}$, we can use the notions of false alarm and miss-detection. A false alarm corresponds to the event that $\mathrm{x}$ predicts outage under hypothesis $H_0$ (i.e., no harmful interference is present); whereas a miss-detection corresponds to the event that $\mathrm{x}$ fails to predict outage under hypothesis $H_1$. Now, the performance of interference model $\mathrm{x}$ can be evaluated using the false alarm and miss-detection probabilities, namely $p_{\mathrm{fa}}^{\mathrm{x}}$ and $p_{\mathrm{md}}^{\mathrm{x}}$. Mathematically speaking,
\begin{align}\label{eq: false-alarm-prob}
p_{\mathrm{fa}}^{\mathrm{x}} &= \Pr \left[\gamma^{\mathrm{x}} < \beta \mid \gamma^{\text{PhyM}} \geq \beta \right] \:, \nonumber \\
p_{\mathrm{md}}^{\mathrm{x}} & = \Pr \left[\gamma^{\mathrm{x}} \geq \beta \mid \gamma^{\text{PhyM}} < \beta \right] \:.
\end{align}

The false alarm and miss-detection probabilities quantify the accuracy of any interference model $\mathrm{x}$ compared to the reference physical model.  Next, we define our accuracy index to be a convex combination of these probabilities.
\begin{defin}[Interference Model Accuracy Index]
For any constant $0 \leq \xi \leq 1$ and   interference model $\mathrm{x}$, the interference model accuracy index is defined as
\begin{align}\label{eq: Definition-IMSindex}
\mathrm{IMA} \left( \mathrm{x}, \xi \right) & = \xi \left( 1 - p_{\mathrm{fa}}^{\mathrm{x} }\right) + \left( 1 - \xi \right) \left( 1 - p_{\mathrm{md}}^{\mathrm{x}}\right) \nonumber \\
& = 1 - \xi \, p_{\mathrm{fa}}^{\mathrm{x} } - \left( 1 - \xi \right) p_{\mathrm{md}}^{\mathrm{x}} \:,
\end{align}
where $p_{\mathrm{fa}}^{\mathrm{x}}$ and $p_{\mathrm{md}}^{\mathrm{x}}$ are given in~\eqref{eq: false-alarm-prob}.
\end{defin}

$\mathrm{IMA} \left( \mathrm{x}, \xi \right)$ is a unit-less real-valued quantity ranging within $[0,1]$, where higher values represent higher similarity between test model $\mathrm{x}$ and the reference physical model.

When ${\xi = \Pr \left[ \gamma^{\text{PhyM}} \geq \beta \right]}$, it follows that $\xi p_{\mathrm{fa}}^{\mathrm{x}} + \left( 1 - \xi \right) p_{\mathrm{md}}^{\mathrm{x}}$ is the average decision error under interference model $\mathrm{x}$. Therefore, $\mathrm{IMA} \left( \mathrm{x}, \Pr \left[\gamma^{\text{PhyM}} \geq \beta \right]\right)$ is the average probability that interference model $\mathrm{x}$ gives the same decision as the reference  physical model.

The proposed index is a universal metric that can be used to quantify the accuracy of different interference models introduced in the literature under different assumptions. We illustrate the use of this index in the next sections, by evaluating the accuracy of IBM and PRM for two example scenarios. For the rest of this paper, we consider $\xi = \Pr \left[ \gamma^{\text{PhyM}} \geq \beta \right]$, so that $\mathrm{IMA} \left( \mathrm{x}, \xi \right)$ evaluates the average probability of correct decision under interference model $\mathrm{x}$.

\section{Scenario~1: Rayleigh Fading Channel with Omnidirectional Communications}\label{sec: Rayleigh-noBlockage}
In this section, we evaluate the accuracy of IBM and PRM for a wireless network exhibiting Rayleigh fading channel and omnidirectional transmission/reception.
%Presence or absence of obstacles on the line-of-sight (LoS) links does not affect the received power at any receiver due to transmission of any transmitter.
These assumptions are, arguably, among the most common assumptions in the analysis and design of wireless networks~\cite{Haenggi2013Stochastic}. Under these assumptions, although PhyM is more tractable for the performance analysis than PRM and IBM~\cite{Haenggi2013Stochastic}, we can derive closed-form expression for the new accuracy index, which in turn results in characterizing its fundamental properties. Nonetheless, even in this network setting, PRM and IBM are more appealing than PhyM for protocol development and for network optimization~\cite{le2010longest}. Moreover, notice that we are not proposing IBM or PRM; rather, we are exemplifying the use of our accuracy index with these well-known interference models.

We consider a reference receiver (called the typical receiver and indexed by 0) at the origin of the Polar coordinate, and its intended transmitter having geometrical/spatial length $d_{00}$. We consider a homogeneous Poisson network of interferers (unintended transmitters) on the plane with density $\lambda_t$ per unit area. For sake of notation simplicity, we drop index 0 from the typical receiver and keep only the indices of the transmitters in~\eqref{eq: protocol-model}. All transmitters are active with transmission power $p$ (no power control). With omnidirectional transmission and reception, there is no antenna gain, so $g_{k}^{t} = g_{k}^{r} = 1$, $k \in \mathcal{I} \cup \{ 0 \}$. Note that we have adopted this set of assumptions to facilitate illustration of the proposed accuracy index, and extension of this paper with more general set of assumptions is straightforward.

Let $\mathcal{B}(\theta,r_{\mathrm{in}},r_{\mathrm{out}})$ be a geometrical annulus sector with angle $\theta$, inner radius $r_{\mathrm{in}}$, and outer radius $r_{\mathrm{out}}$ centered at the location of the typical receiver.
To model a wireless channel, we consider a constant attenuation $a$, distance-dependent attenuation with exponent $\alpha$, and a Rayleigh fading component $h$.
%To avoid the physically unreasonable singularity that arises at the origin under power law attenuation, we modify the path loss index as $\alpha \mathds{1}_{\overline{\mathcal{B}}(2\pi,0,a)}$, where $\mathds{1}_{\cdot}$ is the indicator function taking one over set $\cdot$ and zero otherwise. This modified power law model implies that the signal of all transmitters located outside a disk with radius $a$ will be attenuated by traditional power law method; however, the transmitters inside this disk will observe no channel attenuation.
Therefore, the channel attenuation between transmitter $i$ at radial distance $d_i$ and the typical receiver is $g_{i}^{c} = a h_i d_{i}^{-\alpha}$. We are now ready to illustrate the utility of our interference model accuracy index.

\subsection{Accuracy of the Interference Ball Model}\label{subsec: example-1-IBM}
In this subsection, we derive the accuracy of IBM under the aforementioned system model.
%For mathematical tractability, we assume that $r_{\text{IBM}} \geq a$ and $d_0 \geq a$, and the extension to the general case is straightforward.
We first reformulate the false alarm probability as
\begin{align}\label{eq: false-alarm-uWave-Rayleigh-1}
p_{\mathrm{fa}}^{{\text{IBM}} } &= \Pr \left[\gamma^{\text{IBM}} < \beta \mid \gamma^{\text{PhyM}} \geq \beta \right]  \nonumber \\
& = \frac{\Pr \left[\gamma^{\text{IBM}} < \beta \right] \Pr \left[\gamma^{\text{PhyM}} \geq \beta \mid \gamma^{\text{IBM}} < \beta \right]}{1 - \Pr \left[\gamma^{\text{PhyM}} < \beta \right]} \:.
\end{align}
Although PhyM considers the impacts of all the interferers, IBM considers the effects of the near-field ones. Consequently, $\gamma^{\text{PhyM}} \leq \gamma^{\text{IBM}}$, and thus $\Pr \left[\gamma^{\text{PhyM}} \geq \beta \mid \gamma^{\text{IBM}} < \beta \right]$ in the nominator of~\eqref{eq: false-alarm-uWave-Rayleigh-1} is equal to 0, resulting in $p_{\mathrm{fa}}^{{\text{IBM}}} = 0$.

For the miss-detection probability, we have
\begin{align}\label{eq: miss-detection-uWave-Rayleigh-1}
p_{\mathrm{md}}^{{\text{IBM}} }  &= \Pr \left[\gamma^{\text{IBM}} \geq \beta \mid \gamma^{\text{PhyM}} < \beta \right] \nonumber \\
%& = 1 - \Pr \left[\gamma^{\text{IBM}} < \beta \mid \gamma^{\text{PhyM}} < \beta \right] \nonumber \\
& = 1 - \frac{\Pr \left[\gamma^{\text{IBM}} < \beta \right] \Pr \left[\gamma^{\text{PhyM}} < \beta \mid \gamma^{\text{IBM}} < \beta \right]}{\Pr \left[\gamma^{\text{PhyM}} < \beta \right]} \nonumber \\
& = 1 - \frac{\Pr \left[\gamma^{\text{IBM}} < \beta \right]}{\Pr \left[\gamma^{\text{PhyM}} < \beta \right]} \:,
\end{align}
where the last equality is from $\gamma^{\text{PhyM}} \leq \gamma^{\text{IBM}}$.
Let $\mathbf{E}_{x}$ denote expectation over random variable $x$. Using similar approach as in~\cite{weber2007transmission}, we have
%\begin{figure*}[!t]
%\normalsize
%\begin{align}\label{eq: IBM-1}
%\Pr\left[\gamma^{\text{IBM}} < \beta \right]  = \Pr\left[\frac{p a h_0 d_{0}^{-\alpha}}{\sum\limits_{k \in \mathcal{I} \cap \mathcal{B}(2 \pi, 0, r_{\text{IBM}})} p a h_k d_{k}^{-\alpha} + \sigma}  < \beta \right]
%= 1 - \exp\Vast\{ \frac{- \sigma \beta d_{0}^{\alpha}}{pa} -\pi  \lambda_t  \mathbf{E}_h \vast[ r_{\text{IBM}}^2 \left( 1 - e^{- \beta d_{0}^{\alpha} h r_{\text{IBM}}^{-\alpha}} \right) & \vast. \Vast. \nonumber \\
%\Vast. \vast. + \left( \beta d_{0}^{\alpha} h \right)^{2/\alpha} \Gamma\left( 1 - \frac{2}{\alpha}, \beta d_{0}^{\alpha} h r_{\text{IBM}}^{-\alpha} \right) & \vast] \Vast\}  \:.
%\end{align}
%\hrulefill
%\end{figure*}
\begin{align}\label{eq: IBM-1}
\hspace{-1mm} \Pr\left[\gamma^{\text{IBM}} \hspace{-1mm} <  \hspace{-1mm}  \beta \right] \hspace{-0.5mm} & =
%\Challenge{\hspace{-0.5mm} \Pr\left[\frac{p a h_0 d_{0}^{-\alpha}}{\sum\limits_{k \in \mathcal{I} \cap \mathcal{B}(2 \pi, 0, r_{\text{IBM}})} p a h_k d_{k}^{-\alpha} + \sigma}  < \beta \right]} \nonumber \\
\hspace{-0.5mm} 1 \hspace{-0.8mm} - \hspace{-0.6mm} \exp \hspace{-0.8mm} \vast\{ \hspace{-1mm} \frac{- \sigma \beta d_{0}^{\alpha}}{pa} \hspace{-0.3mm} - \hspace{-0.3mm} \pi  \lambda_t  \mathbf{E}_h \hspace{-0.8mm} \Bigg[ r_{\text{IBM}}^2 \hspace{-0.5mm} \left( \hspace{-0.5mm} 1 \hspace{-0.8mm} - \hspace{-0.5mm} e^{- \beta d_{0}^{\alpha} h r_{\text{IBM}}^{-\alpha}} \hspace{-0.4mm} \right) \vast.  \nonumber \\[-3mm]
&  \hspace{7mm} \vast. + \left( \beta d_{0}^{\alpha} h \right)^{2/\alpha} \Gamma \hspace{-0.3mm} \left( \hspace{-0.2mm} 1 - \frac{2}{\alpha}, \beta d_{0}^{\alpha} h r_{\text{IBM}}^{-\alpha} \hspace{-0.2mm} \right) \hspace{-0.7mm} \Bigg] \hspace{-0.7mm} \vast\} , \hspace{-4mm}
\end{align}
where $\Gamma\left(\cdot, \cdot \right)$ is the incomplete Gamma function, and probability density function of $h$ is $f_h(x) = e^{-x}$. Detailed derivation steps are avoided from this paper due to space limitation, but are presented in the extended version~\cite{Shokri2015IMSindex}. Proofs of the following results can be also found in~\cite{Shokri2015IMSindex}.
To find $\Pr \left[\gamma^{\text{PhyM}} < \beta \right]$, we evaluate $\Pr \left[\gamma^{\text{IBM}} < \beta \right]$ at $r_{\text{IBM}} \to \infty$. Therefore, $\Pr \left[\gamma^{\text{PhyM}} < \beta \right]$ is equal to
\begin{align}\label{eq: PhyM-1}
1 - \exp \hspace{-0.3mm} \vast\{ \hspace{-1.5mm} - \frac{\sigma \beta d_{0}^{\alpha}}{pa} - \pi  \lambda_t \mathbf{E}_h \hspace{-0.4mm} \Biggl[ \hspace{-0.8mm} \left( \beta d_{0}^{\alpha} h \right)^{\frac{2}{\alpha}}\Gamma\hspace{-0.4mm} \left( \hspace{-0.5mm} 1 - \frac{2}{\alpha} \hspace{-0.3mm} \right)\hspace{-0.7mm} \Biggr] \hspace{-0.7mm} \vast\} \:,
\end{align}
where $\Gamma\left(\cdot \right)$ is the Gamma function.
%\begin{align}\label{eq: PhyM-1}
%\Pr \left[\gamma^{\text{PhyM}} < \beta \right] \hspace{-0.5mm} & = \Challenge{\hspace{-0.5mm} \lim_{r_{\text{IBM}} \to \infty} \, \Pr\left[\gamma^{\text{IBM}} < \beta \right] } \nonumber \\
%& = \hspace{-0.5mm} 1 \hspace{-0.5mm} - \hspace{-0.5mm} \exp \hspace{-0.5mm} \Vast\{ \hspace{-1.7mm} - \frac{\sigma \beta d_{0}^{\alpha}}{pa} \vast. \Vast. \nonumber \\
%& \hspace{16.2mm} \Vast. \vast. - \pi  \lambda_t \mathbf{E}_h \hspace{-0.8mm} \vast[ \hspace{-0.8mm} \left( \beta d_{0}^{\alpha} h \right)^{\frac{2}{\alpha}}\Gamma\hspace{-0.5mm} \left( \hspace{-0.5mm} 1 - \frac{2}{\alpha} \hspace{-0.3mm} \right)\hspace{-0.8mm} \vast] \hspace{-0.8mm} \Vast\} \:.
%\end{align}
Substituting~\eqref{eq: IBM-1}--\eqref{eq: PhyM-1} into~\eqref{eq: miss-detection-uWave-Rayleigh-1}, the miss-detection probability follows. Also, from~\eqref{eq: Definition-IMSindex}, the accuracy of the interference ball model $\mathrm{IMA} \left( \text{IBM}, \Pr \left[ \gamma^{\text{PhyM}} \geq \beta \right] \right)$ is derived.

\begin{result}[Perfect Interference Ball Model]\label{prop: perfect-IBM}
For any constant $0 \leq \xi \leq 1$, $\mathrm{IMA} \left( \text{IBM}, \xi \right) \to 1$ as $r_{\text{IBM}} \to \infty$.
\end{result}

%\begin{IEEEproof}
%We know that $p_{\mathrm{fa}}^{{\text{IBM}}} = 0$ for any $0 \leq \xi \leq 1$. Moreover, as $r_{\text{IBM}}$ increases, $\Pr \left[\gamma^{\text{IBM}} < \beta \right]$ becomes closer to $\Pr \left[\gamma^{\text{PhyM}} < \beta \right]$. Considering~\eqref{eq: miss-detection-uWave-Rayleigh-1}, $P_{\mathrm{md}}^{{\text{IBM}} }$ asymptotically goes to zero as $r_{\text{IBM}} \to \infty$. With zero false alarm and asymptotically zero miss-detection probabilities, the proof is concluded from~\eqref{eq: Definition-IMSindex}.
%\end{IEEEproof}
Result~\ref{prop: perfect-IBM} implies that the accuracy of IBM increases with $r_{\text{IBM}}$, and it can be arbitrary accurate for sufficiently large $r_{\text{IBM}}$. The price, however, is more complicated IBM for the protocol development and network optimization~\cite{le2010longest} as it accounts for more interferers. Also, negotiation with other transmitters (e.g., for joint power control or scheduling) within $r_{\text{IBM}}$ becomes more complicated.

\subsection{Accuracy of the Protocol Model}\label{subsec: example-1-PRM}
We now consider the PRM and first note that
\begin{align}\label{eq: false-alarm-prop2}
p_{\mathrm{fa}}^{{\text{PRM}} } \hspace{-0.6mm} &= \hspace{-0.6mm} \Pr \left[\gamma^{\text{PRM}} < \beta \mid \gamma^{\text{PhyM}} \geq \beta \right] \nonumber \\
%& = \hspace{-0.6mm} 1 \hspace{-0.5mm} - \hspace{-0.5mm} \Pr \left[\gamma^{\text{PRM}} \geq \beta \mid \gamma^{\text{PhyM}} \geq \beta \right] \nonumber \\
& = \hspace{-0.6mm} 1 \hspace{-0.5mm} - \hspace{-0.5mm} \frac{\hspace{-0.4mm} \left( \hspace{-0.3mm} 1 \hspace{-0.5mm} - \hspace{-0.5mm}  \Pr \left[\gamma^{\text{PRM}} \hspace{-0.4mm} < \hspace{-0.4mm} \beta \right] \right) \hspace{-1mm} \left( 1 \hspace{-0.5mm} - \hspace{-0.5mm} \Pr \left[\gamma^{\text{PhyM}} \hspace{-0.4mm} < \hspace{-0.4mm} \beta \hspace{-0.7mm} \mid  \hspace{-0.7mm} \gamma^{\text{PRM}} \hspace{-0.4mm}  \geq  \hspace{-0.4mm} \beta \right]\right)\hspace{-0.4mm}}{1 - \Pr \left[\gamma^{\text{PhyM}} < \beta \right]} .
\end{align}
and that
\begin{align}\label{eq: miss-detection-prop-2}
p_{\mathrm{md}}^{{\text{PRM}} } &= \Pr \left[\gamma^{\text{PRM}} \geq \beta \mid \gamma^{\text{PhyM}} < \beta \right] \nonumber \\
& = \frac{ \left(1 -  \Pr \left[\gamma^{\text{PRM}} < \beta \right] \right) \Pr \left[\gamma^{\text{PhyM}} < \beta \mid \gamma^{\text{PRM}} \geq \beta \right]}{\Pr \left[\gamma^{\text{PhyM}} < \beta \right]} \:.
\end{align}
In the last two equations, note that $\Pr[\gamma^{\text{PhyM}} < \beta ]$ is derived in~\eqref{eq: PhyM-1}. In the following, we evaluate $\Pr[\gamma^{\text{PRM}} < \beta]$ and $\Pr[\gamma^{\text{PhyM}} < \beta \mid \gamma^{\text{PRM}} \geq \beta]$.

Event $\gamma^{\text{PRM}} < \beta$ occurs if there is at least one interferer inside $\mathcal{B}(2\pi,0,r_{\text{PRM}})$. As $\mathcal{I}$ is a homogenous Poisson point process with intensity $ \lambda_t$, we have
\begin{equation}\label{eq: PRM-1}
\Pr \left[\gamma^{\text{PRM}} < \beta \right] = 1 - \exp\Big\{\hspace{-0.7mm} - \lambda_t \pi r_{\text{PRM}}^2 \Big\} \:.
\end{equation}
Event $\gamma^{\text{PRM}} \geq \beta$ implies that there is no interferer inside $\mathcal{B}(2\pi,0,r_{\text{PRM}})$. Therefore, $\Pr[\gamma^{\text{PhyM}} < \beta \mid \gamma^{\text{PRM}} \geq \beta]$ is given in~\eqref{eq: PRM-2} on the top of page~\pageref{eq: PRM-2}, where $\mathds{1}_{\cdot}$ is the indicator function taking one over set $\cdot$ and zero otherwise.
\begin{figure*}[!t]
\normalsize
\begin{align}\label{eq: PRM-2}
\hspace{-1.5mm} 1 - \exp \hspace{-0.5mm} \vast\{ \hspace{-0.8mm} \frac{- \sigma \beta d_{0}^{\alpha}}{pa} - \pi  \lambda_t  \mathbf{E}_h \hspace{-0.8mm} \Bigg[ \hspace{-0.8mm} - r_{\text{PRM}}^2 \left( 1 - e^{- \beta d_{0}^{\alpha} h r_{\text{PRM}}^{-\alpha}} \right) + \left( \beta d_{0}^{\alpha} h \right)^{2/\alpha} \Gamma\hspace{-0.8mm}\left(\hspace{-0.4mm} 1 - \frac{2}{\alpha} \hspace{-0.4mm} \right) - \left( \beta d_{0}^{\alpha} h \right)^{2/\alpha} \Gamma \hspace{-0.8mm} \left( \hspace{-0.4mm} 1 - \frac{2}{\alpha}, \beta d_{0}^{\alpha} h r_{\text{PRM}}^{-\alpha} \hspace{-0.4mm} \right)\hspace{-0.8mm}  \Bigg] \hspace{-0.8mm} \vast\}  .
\end{align}
\hrulefill
\vspace{-2mm}
\end{figure*}

%\end{figure*}
%\begin{figure*}[!t]
%\normalsize
%\begin{align}\label{eq: PRM-2}
%\Pr[\gamma^{\text{PhyM}} < \beta \mid \gamma^{\text{PRM}} \geq \beta]  &= \Challenge{1 - \exp\Vast\{\frac{- \sigma \beta d_{0}^{\alpha}}{pa} -2 \pi  \lambda_t  \mathbf{E}_h \vast[\int_{0}^{\infty} \! \mathds{1}_{\mathcal{B}(2\pi,r_{\text{PRM}},\infty)} \left( 1 - e^{- \beta d_{0}^{\alpha} h r^{-\alpha}} \right) r \, \mathrm{d}r \vast] \Vast\} }\nonumber \\
%&= 1 - \exp\Vast\{ \frac{- \sigma \beta d_{0}^{\alpha}}{pa} - \pi  \lambda_t  \mathbf{E}_h \vast[ - r_{\text{PRM}}^2 \left( 1 - e^{- \beta d_{0}^{\alpha} h r_{\text{PRM}}^{-\alpha}} \right) + \left( \beta d_{0}^{\alpha} h \right)^{2/\alpha} \Gamma\left( 1 - \frac{2}{\alpha} \right) \vast. \Vast. \nonumber \\
%& \hspace{70mm} \Vast. \vast.
%- \left( \beta d_{0}^{\alpha} h \right)^{2/\alpha} \Gamma\left( 1 - \frac{2}{\alpha}, \beta d_{0}^{\alpha} h r_{\text{PRM}}^{-\alpha} \right) \vast] \Vast\}  \:.
%\end{align}
%\hrulefill
%\end{figure*}
Substituting~\eqref{eq: PhyM-1} and \eqref{eq: PRM-1}--\eqref{eq: PRM-2} into \eqref{eq: false-alarm-prop2}--\eqref{eq: miss-detection-prop-2}, and the results into~\eqref{eq: Definition-IMSindex}, we can find $\mathrm{IMA} \left( \text{PRM}, \Pr \left[ \gamma^{\text{PhyM}} \geq \beta \right] \right)$.

\begin{result}[Miss-detection--False Alarm Tradeoff]\label{result: Pmd-Pfa-tradeoff}
Consider the protocol model of interference with Rayleigh fading channel. Increasing the interference range $r_{\text{PRM}}$ reduces the false alarm probability and increases the miss-detection probability. Decreasing the interference range increases the false alarm probability and reduces the miss-detection probability.
\end{result}

\section{Scenario~2: Deterministic Channel, Directional Communications, with Obstacles}\label{sec: Rayleigh-Blockage}
In this section, we investigate the accuracy of the IBM and PRM in modeling a wireless network with deterministic channel condition, directional communications, and impenetrable obstacles in the environment. Application areas include modeling and performance evaluation of mmWave networks, where the sparse scattering characteristic of the mmWave frequencies and the narrow-beam operation make the mmWave channel more deterministic compared to that of traditional microwave systems having rich scattering environment and omnidirectional communication~\cite{Rangan2014Millimeter}. Moreover, extreme penetration loss in the mmWave frequencies (e.g., 35~dB due to the human body~\cite{Rangan2014Millimeter}) justifies the assumption of impenetrable obstacles in Scenario~2.

We assume similar homogenous Poisson network of interferers as in Section~\ref{sec: Rayleigh-noBlockage}. If there is no obstacle on the link between transmitter $i$ and the typical receiver, we say that transmitter $i$ has line-of-sight (LoS) condition with respect to the typical receiver, otherwise it is in non-LoS condition. We assume that transmitter of every link is spatially aligned with its intended receiver, so there is no beam-searching phase. The effects of beam-searching phase is analyzed in~\cite{Shokri2015Beam}.
We assume the same operating beamwidth $\theta$ for all devices in both transmission and reception modes. Motivated by the large number of antenna elements in mmWave systems and for mathematical tractability, we neglect the sidelobe radiations from all interference models (PRM, IBM, and PhyM). Moreover, we model the antenna pattern with an ideal sector model, where the antenna gain for each transmitter/receiver is $2 \pi /\theta$ in the main lobe~\cite{Shokri2015Beam}.
%At the MAC layer, the beamforming can be represented by an ideal sector antenna pattern~\cite{di2014stochastic}, where the antenna gain is a constant for all angles in the main lobe and equal to a smaller constant in the side lobe.
%%These gains mainly depend on the number of antenna elements and their configuration.
%Motivated by the large number of antenna elements in mmWave systems and for mathematical tractability, we neglect the sidelobe radiations from all interference models (PRM, IBM, and PhyM). Assuming the same operating beamwidth $\theta$ for all devices in both transmission and reception modes, neglecting the sidelobe radiations, and considering a 2D beamforming, we have that the antenna gain for each transmitter/receiver is $2 \pi /\theta$ on the main lobe~\cite{Shokri2015Beam}.
With a random number of obstacles, each having random location and size, we see that the link between transmitter $i$ and receiver $j$ with length $d_{ij}$ is in the LoS condition with probability $e^{-\epsilon \lambda_o d_{ij}}$, where $\lambda_o$ is the density of obstacles per unit area and $\epsilon$ is a constant value that depends on the average size of the obstacles in the environment~\cite{TBai2014Blockage}. Due to the exponential decrease of the LoS probability with the link length, very far transmitters are most likely blocked. As in~\cite{TBai2014Blockage,di2014stochastic}, we assume independent LoS conditions among the typical receiver and all other transmitters. Again, we are using this system model to highlight the fundamental properties of the accuracy index. The exact value of this index can be easily numerically calculated under any system model, not just the one considered in this section.

To evaluate the accuracy of the IBM and PRM, we first note that an interferer can give a significant interference contribution at the typical receiver if: (a) the typical receiver is inside its main lobe, (b) it has LoS condition with respect to the typical receiver, and (c) it is inside the main lobe of the typical receiver.
Due to random deployment of the transmitters/receivers, the probability that the typical receiver locates inside the main lobe of a transmitter is $\theta/2 \pi$. Moreover, we have independent LoS events among the typical receiver and individual transmitters. Therefore, if the transmitter density per unit area is $\lambda_t$, the interferers for which conditions~(a)--(b) hold follow an inhomogeneous Poisson point process with intensity of ${\lambda_I \left( r \right) =  \lambda_t \theta e^{-\epsilon\lambda_o r} / 2\pi}$ at radial distance $r$.
%We note that $\mathcal{I} \cap \mathcal{B}(\theta,0,r_{\text{PRM}})$ is the set of potential interferers inside the vulnerable region of the protocol model, shown by red triangles in Fig.~\ref{fig: IntRegion}, and $\mathcal{I} \cap \mathcal{B}(\theta,r_{\text{PRM}},\infty)$ shows the set of potential interferers outside that region, shown by green circles in Fig.~\ref{fig: IntRegion}. Also, $\mathcal{I} \cap \mathcal{B}(\theta,0,r_{\text{IBM}})$ is the set of potential interferers for IBM (near-field interferers).
%\begin{figure}[!t]
%  \centering
%  \includegraphics[width=0.6\columnwidth]{Figures/IntRegion}\\
%
%  \caption{Illustration of the vulnerable area.}\label{fig: IntRegion}
%\end{figure}
%\subsection{Impact of Directionality and Blockage}
In the following, we investigate the impacts of directionality and blockage on the accuracy of the interference models. We define by $\Lambda_{\mathcal{B}(\theta,0,R)}$ the measure of region $\mathcal{B}(\theta,0,R)$, i.e., the average number of interferers inside the region. Thus,
\begin{equation}\label{eq: measure-of-region}
\Lambda_{\mathcal{B}(\theta,0,R)}\hspace{-0.3mm}  = \hspace{-0.3mm}  \theta \hspace{-1.5mm} \int_{0}^{R}\hspace{-2mm}  {\! \lambda_I \left( r \right) r\, \mathrm{d}r}\hspace{-0.5mm} = \hspace{-0.5mm} \frac{\theta^2  \lambda_t}{2\pi \epsilon^{2} \lambda_{o}^{2}} \Big( 1 - \left( 1 + \epsilon \lambda_o R \right)e^{- \epsilon \lambda_o R}\Big) \:.
\end{equation}
%Then, for any real $R>0$, the number of potential interferer inside the region $\mathcal{B}(\theta,0,R)$, denoted by $N_{\mathcal{B}(\theta,0,R)}$, is a Poisson random variable with probability mass function
%\begin{equation}\label{eq: number-of-points}
%  \Pr[N_{\mathcal{B}(\theta,0,R)}=n] = e^{-\Lambda_{\mathcal{B}(\theta,0,R)}} \frac{\left(\Lambda_{\mathcal{B}(\theta,0,R)} \right)^{n}}{n!} \:.
%\end{equation}

\begin{result}[Impact of Directionality and Blockage]\label{result: impact-of-obstacle}
Consider~\eqref{eq: measure-of-region}, and let ${R \to \infty}$. The average number of potential interferers converges to
\begin{equation}\label{eq: avg-number-interferers}
\frac{\theta^2  \lambda_t}{2\pi \epsilon^{2} \lambda_{o}^{2}} \:,
\end{equation}
which does not diverge almost surely if $\epsilon\lambda_o > 0$.
\end{result}

%To interpret Result~\ref{result: impact-of-directionality}, with no obstacle in the environment (${\epsilon\lambda_o \to 0}$), we will have a homogenous Poisson network of interferers with density $ \lambda_t \theta/ 2 \pi$. Therefore, the average number of interferers over $\mathcal{B}(\theta,0,R)$ is the product of the density per unit area and the area of $\mathcal{B}$, which is $\theta R^2/2$.
%From Result~\ref{result: impact-of-directionality}, adopting narrower beams reduces the average number of potential interferers within a certain distance $R$; however, it still tends to infinity almost surely as $R \to \infty$.

Result~\ref{result: impact-of-obstacle} implies that any receiver observes a finite number of potential interferers almost surely if there is a non-negligible blockage.\footnote{In the conventional microwave systems where the transmission is less sensitive to blockage, the number of potential interferers is almost surely infinite~\cite{Haenggi2013Stochastic}.} This unique feature holds for the mmWave bands,as most of the obstacles can severely attenuate the mmWave signals. Therefore, not only farther transmitters will contribute less on the aggregated interference (due to higher distance-dependent path-loss) but they will be also thinned by directionality and blockage such that only a \emph{finite} number of spatially close transmitters can cause non-negligible interference to any receiver. This indeed makes the physical model of interference closer to IBM, which considers only the near-field interferers. %To elaborate more, we present the following result.

%{\color{red}\begin{result}[Measure of Far-Field Interferers]\label{prop: measure-far-field}
%The average number of interferers located inside $\mathcal{B}(\theta,R,\infty)$ is
%\begin{equation}\label{eq: measure-far-field}
%\Lambda_{\mathcal{B}(\theta,R,\infty)} = \frac{\theta^2  \lambda_t}{2\pi \epsilon^{2} \lambda_{o}^{2}} \left( 1 + \epsilon \lambda_o R \right)e^{- \epsilon \lambda_o R} \:,
%\end{equation}
%and the probability of having at least one far-field interferer is
%\begin{equation}\label{eq: AvgNoPoints-far-field}
%1 - e^{-\Lambda_{\mathcal{B}(\theta,R,\infty)}} \:.
%\end{equation}
%\end{result}
%}
%\begin{IEEEproof}
%The proof is straightforward and omitted from this paper.
%\end{IEEEproof}

%{\color{red}From Result~\ref{prop: measure-far-field}, the probability of having no far-field interferers increases exponentially with the distance. By defining any arbitrary minimum threshold $\kappa$ for~\eqref{eq: AvgNoPoints-far-field}, we can find a distance $R_{\kappa}$ after which the probability of having far-field interferer(s) is arbitrary close to 0 (less than $\kappa$). This suggests that by setting $r_{\text{IBM}} = R_{\kappa}$, the IBM can capture, at least, $(1-\kappa)$ fraction of the total number of interferers for any arbitrary small $\kappa$. Recall that the neglected interferers, if any, are far-field, and their contributions on the total interference term are suppressed by the significant distance-dependent path-loss.} All these facts result in the following conclusion:
\begin{result}
Directionality and blockage can increase the accuracy of the interference ball model.
\end{result}
We can show similar accuracy improvement in the PRM, as we numerically illustrate in the next section.
%\begin{figure}[!t]
%  \centering
%  \includegraphics[width=\columnwidth]{Figures/FarFieldInterference}\\
%  \caption{Probability of having at least one far-field interferer as a function of distance.}\label{fig: FarFieldInterference}
%\end{figure}

\subsection{Accuracy of the Interference Ball Model}\label{subsec: example-2-IBM}
Considering Section~\ref{subsec: example-1-IBM}, we immediately see that ${p_{\mathrm{fa}}^{{\text{IBM}}} = 0}$. Moreover, we have that, for any $0 \leq \xi \leq 1$, $\mathrm{IMA} \left( \text{IBM}, \xi \right) \to 1$ as $r_{\text{IBM}} \to \infty$. However, the miss-detection probability, and consequently $\mathrm{IMA} \left( \text{IBM}, \xi \right)$, cannot be derived in general in a tractable closed-form expression. In the extended version of this paper~\cite{Shokri2015IMSindex}, we have derived an upper bound for the miss-detection probability, for which we substitute a lower bound of $\gamma^{\text{IBM}}$ and an upper bound of $\gamma^{\text{PhyM}}$ into~\eqref{eq: miss-detection-uWave-Rayleigh-1}. In the next section, we will numerically evaluate $\mathrm{IMA} \left( \text{IBM}, \Pr \left[ \gamma^{\text{PhyM}} \geq \beta \right] \right)$.

\subsection{Accuracy of the Protocol Model}\label{subsec: example-2-PRM}
Again, we cannot find tractable closed-form expressions for the false alarm and miss-detection probabilities with deterministic wireless channel. Nonetheless, we can characterize some properties of the accuracy index for the protocol model.
We first observe that Result~\ref{result: Pmd-Pfa-tradeoff} holds here. Moreover, we have the following result.
\begin{result}[Zero False Alarm Probability]\label{result: zero-false-alarm-IV}
Consider the deterministic channel model. The false alarm probability is zero for any $r_{\text{PRM}} \leq \zeta^{-1/\alpha}$, where
\begin{equation}\label{eq: zeta}
\zeta =   \frac{d_{0}^{-\alpha}}{\beta} - \frac{\sigma}{pa}\left( \frac{\theta }{2 \pi} \right)^{2} \:.
\end{equation}
\end{result}
As we discussed in~\cite{Shokri2015IMSindex}, the zero false alarm probability is a consequence of the deterministic channel model. In the following, we will numerically illustrate the accuracy index as well as Results~\ref{prop: perfect-IBM}--\ref{result: zero-false-alarm-IV}.

\section{Numerical Results}\label{sec: numerical-results}
We simulate a spatial Poisson network of interferers and obstacles with density $\lambda_t$ and $\lambda_o$ per unit area. Length of the typical link is 20~m. For Scenario~1 (Section~\ref{sec: Rayleigh-noBlockage}), we simulate a traditional outdoor microwave network~\cite{di2014stochastic} with average attenuation $a=22.7$~dB at the reference distance 1~m, path-loss index $\alpha=3.6$, and noise power $\sigma = -111$~dBm (around 2~MHz bandwidth). For Scenario~2 (Section~\ref{sec: Rayleigh-Blockage}), we simulate a mmWave network at 28~GHz~\cite{di2014stochastic} with $a=-61.4$~dB, $\alpha=2.5$, $\sigma = -81$~dBm (around 2~GHz bandwidth), and $\epsilon \lambda_o = 0.008$~\cite{TBai2014Blockage}.
For both scenarios, we consider $p=20$~dBm transmission power and $\beta = 5$~dB minimum SINR threshold. For the ease of illustration, we define the notion of the \emph{average inter-transmitter distance} as $d_t = 1/\sqrt{\lambda_t}$.
%This distance directly relates to the inter-site distance in cellular networks, and also shows the transmitter density in a network.

\begin{figure}[!t]
  \centering
  \subfigure[]{
    \centering
    \includegraphics[width=\columnwidth]{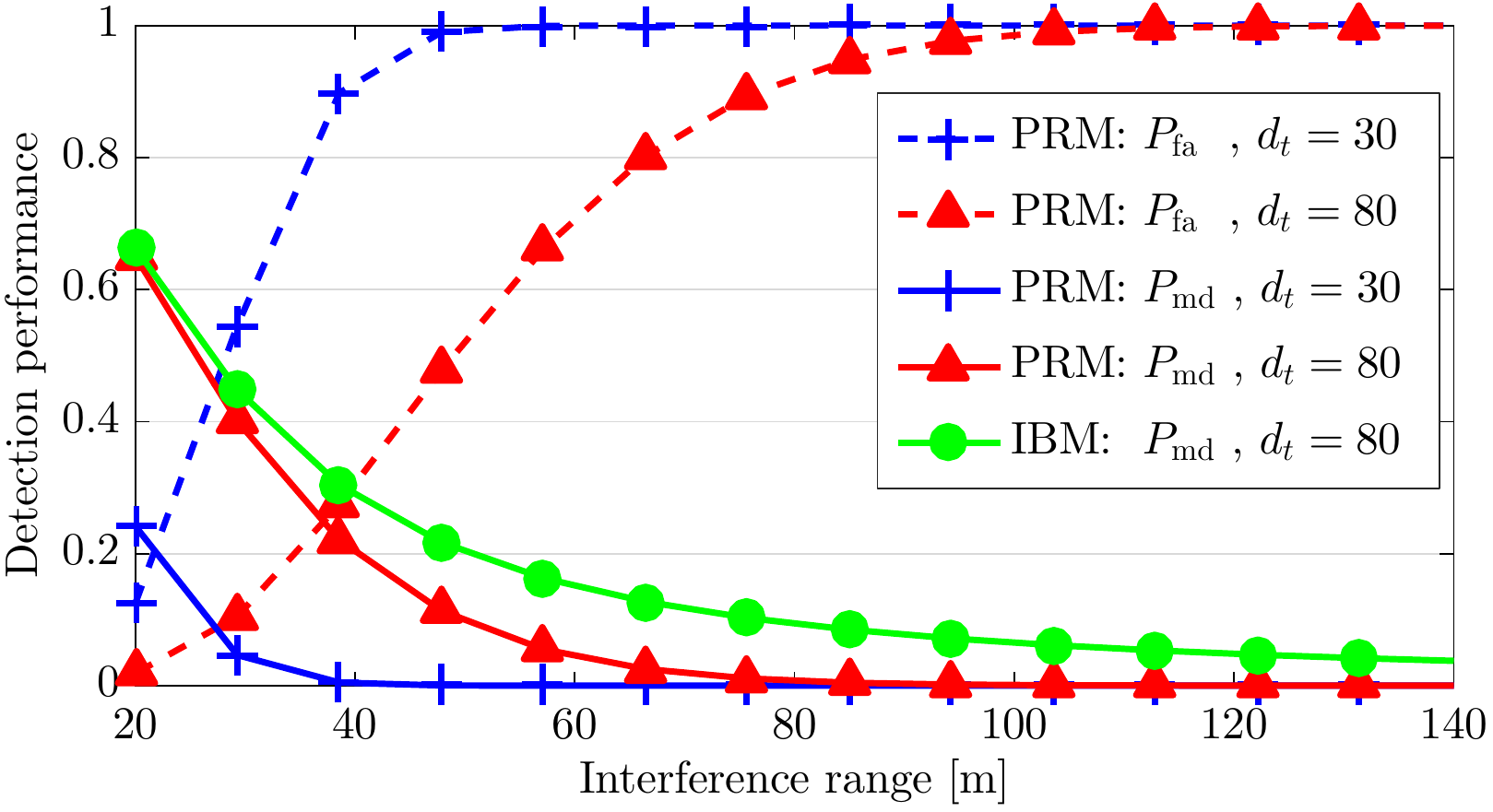}
    \label{subfig: uWave_Omni_Rayleigh__r_PfaPmd}
  }
  \subfigure[]{
       \centering
%    \vspace{4.1mm}
  \includegraphics[width=\columnwidth]{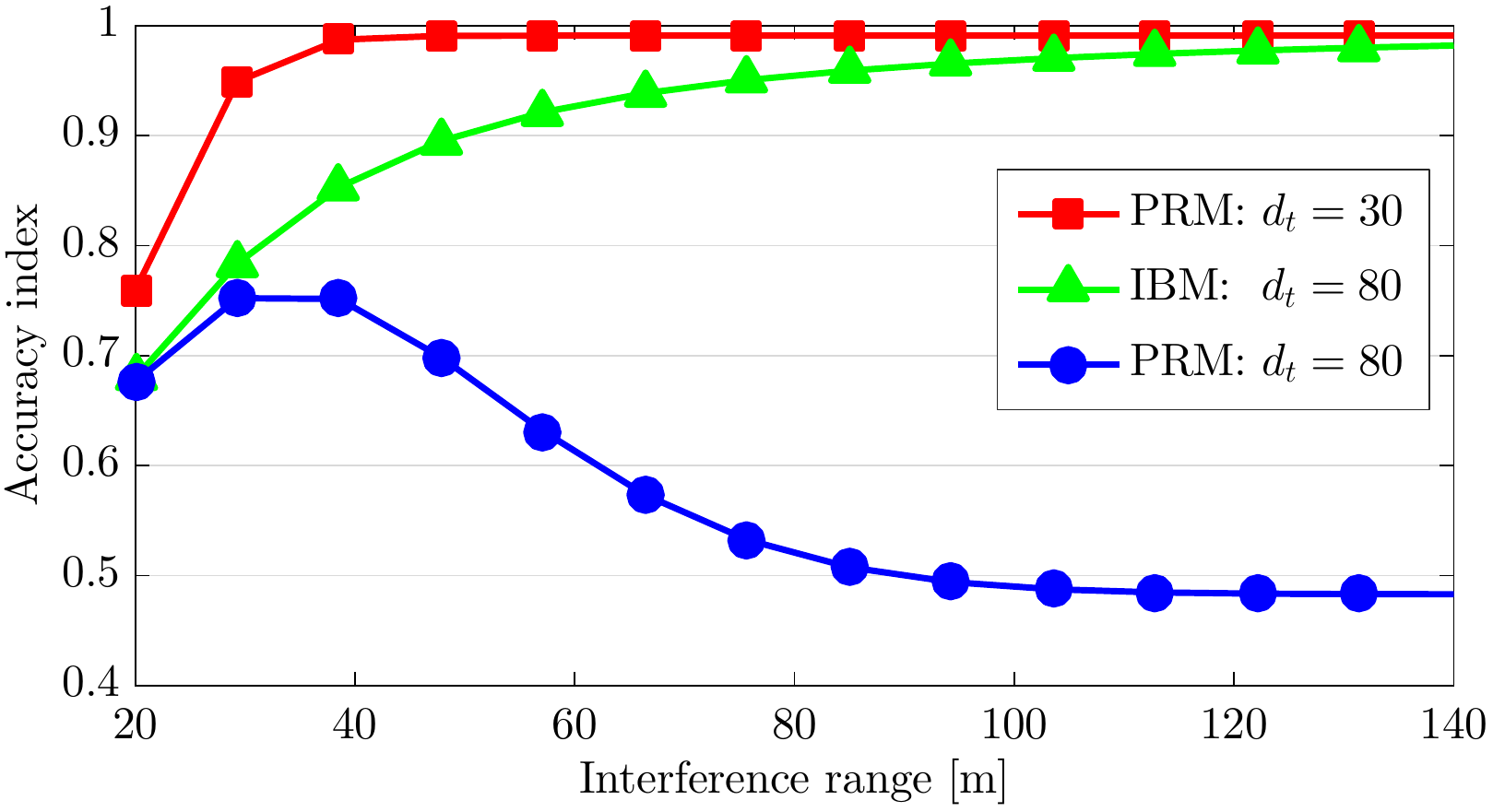}
    \label{subfig: uWave_Omni_Rayleigh__r_IMA}
    }
  \caption{Impact of the interference range on the accuracy of interference models under Rayleigh fading channel and omnidirectional communications.}
  \label{fig: uWave_Omni_Rayleigh__r}
\vspace{-2mm}
\end{figure}
Fig.~\ref{fig: uWave_Omni_Rayleigh__r} illustrates the impact of the interference range on the accuracy of both IBM and PRM under Scenario~1. From Fig.~\ref{subfig: uWave_Omni_Rayleigh__r_PfaPmd}, increasing $r_{\text{PRM}}$ increases $p_{\mathrm{fa}}^{{\text{PRM}}}$ and reduces $p_{\mathrm{md}}^{{\text{PRM}}}$, highlighted as the tradeoff between the miss-detection and false alarm probabilities, stated in Result~\ref{result: Pmd-Pfa-tradeoff}. This tradeoff may lead to increment (see $d_t = 30$) or decrement (see $d_t = 80$) of the accuracy index of the PRM with the interference range. The IBM has zero false alarm probability, not depicted in Fig.~\ref{subfig: uWave_Omni_Rayleigh__r_PfaPmd} for sake of clarity of the figure. Moreover, as stated in Result~\ref{prop: perfect-IBM}, $p_{\mathrm{md}}^{{\text{IBM}}}$ decreases with $r_{\text{PRM}}$, leading to a more accurate IBM, as can be confirmed in Fig.~\ref{subfig: uWave_Omni_Rayleigh__r_IMA}. Note that with the same transmitter density and interference range, the PRM has lower miss-detection probability than the IBM; however, better false alarm performance of the IBM leads to less errors in detecting outage events and therefore higher accuracy index.

\begin{figure}[!t]
	\centering
    \subfigure[Interference ball model]{
       \centering
		\includegraphics[width=\columnwidth]{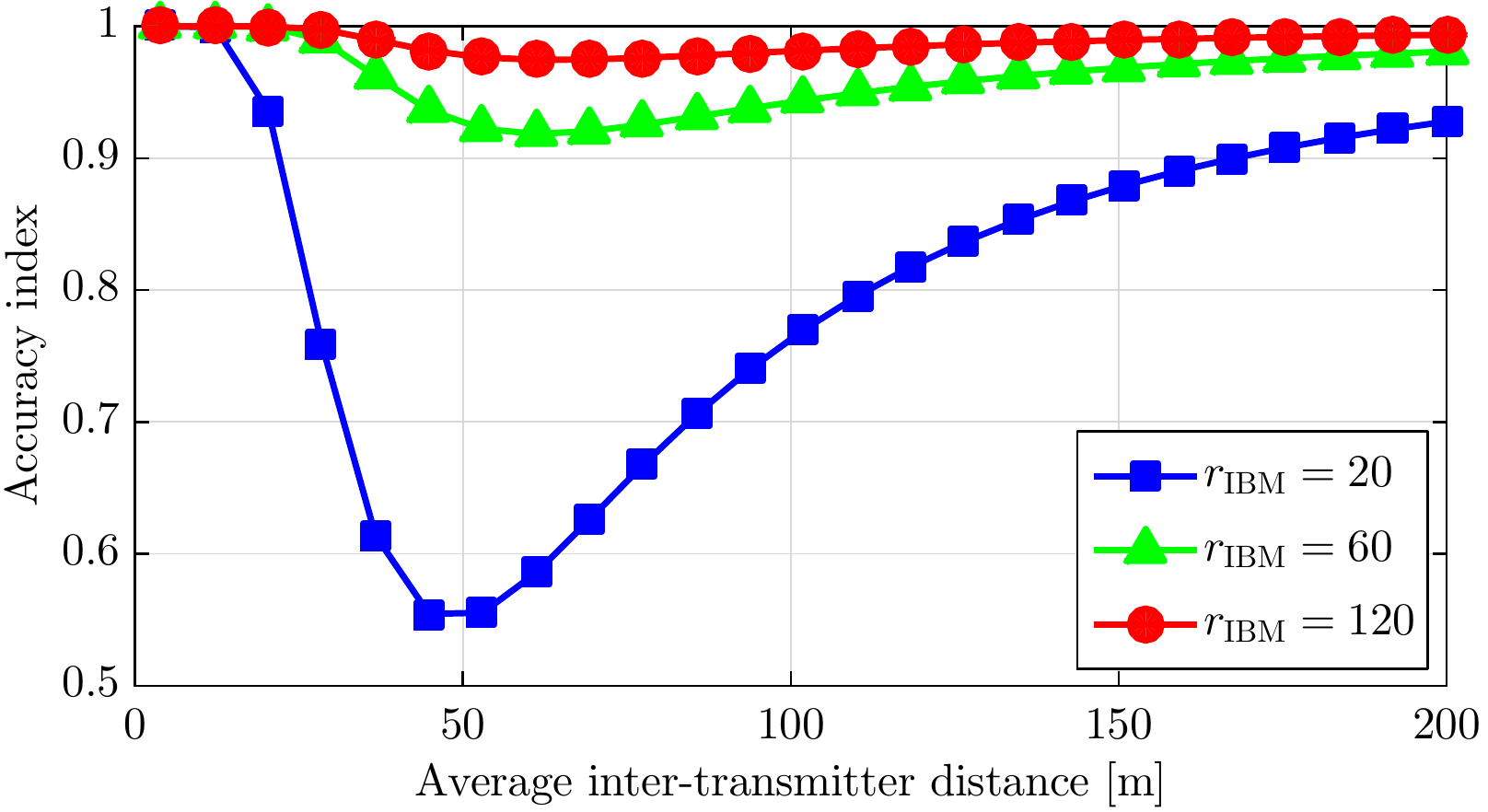}
	   \label{subfig: IBM_uWave_Omni_Rayleigh__lambdaT}
    }
    \subfigure[Protocol model of interference]{
       \centering
%       \vspace{5mm}
  \includegraphics[width=\columnwidth]{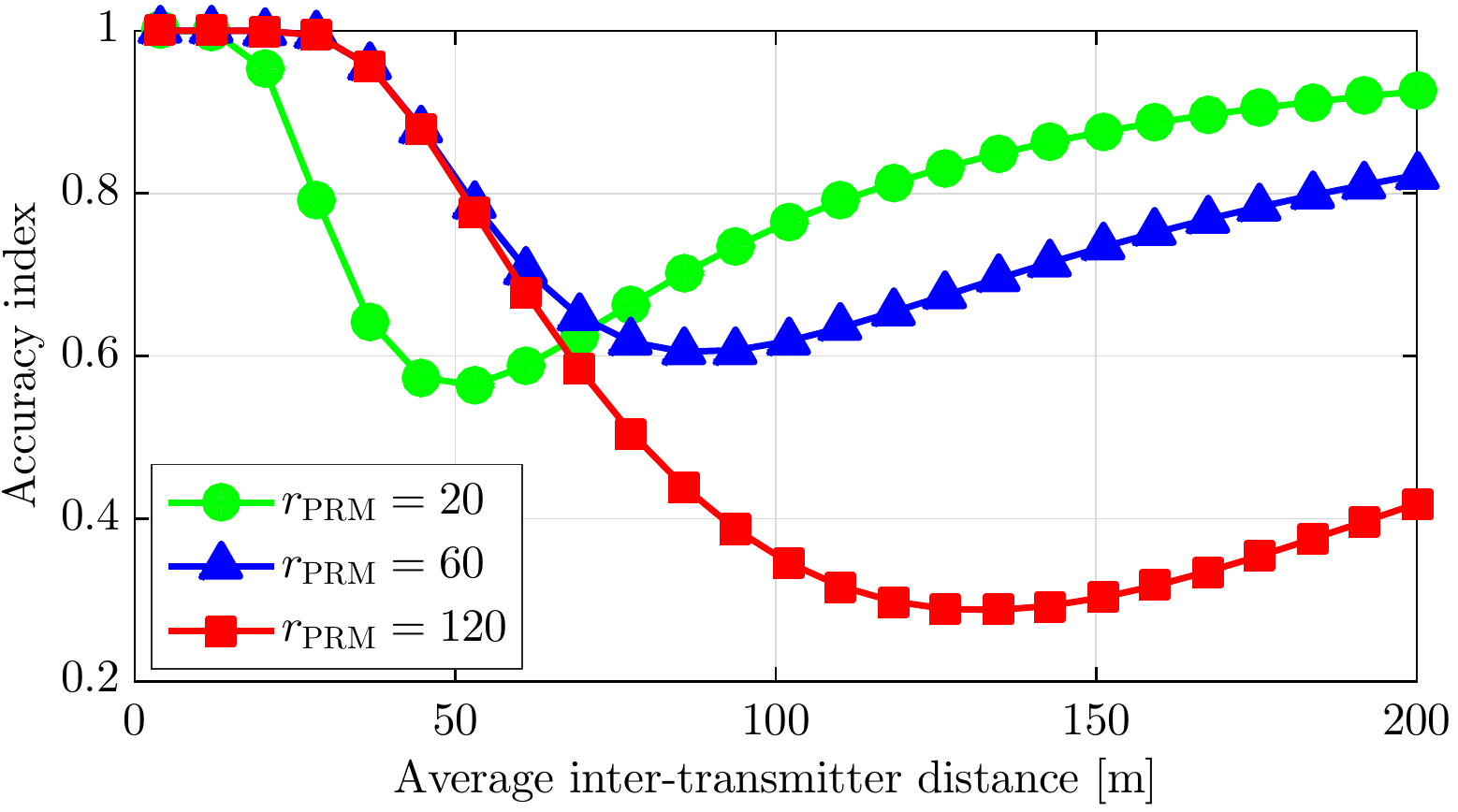}
       \label{subfig: PRM_uWave_Omni_Rayleigh__lambdaT}
    }		
    \caption{Impact of transmitter density on the accuracy of the interference models under Rayleigh fading channel and omnidirectional communications.}
	\label{fig: uWave_Omni_Rayleigh__lambdaT}
\vspace{-2mm}
\end{figure}

Fig.~\ref{fig: uWave_Omni_Rayleigh__lambdaT} shows the accuracy of the IBM and PRM under Scenario~1 against the average inter-transmitter distance $d_t$. Again, we can observe enhancement in the accuracy of the IBM with $r_{\text{IBM}}$, whereas the accuracy index of the PRM shows a complicated behavior as a function of $r_{\text{PRM}}$. Both interference models are very accurate at extremely dense transmitter deployments. In fact, the interference level is so high in this case that $\xi = \Pr \left[ \gamma^{\text{PhyM}} \geq \beta \right]$ is almost 0, and therefore the accuracy index is determined only by the miss-detection probability. And, increasing the transmitter density (lower $d_t$) decreases this probability for both IBM and PRM, see Fig.~\ref{subfig: uWave_Omni_Rayleigh__r_PfaPmd}, improving their accuracy. For ultra sparse transmitter deployments, again, both interference models work accurately, as $\xi \to 1$ in this case and therefore only the false alarm probability will determine the accuracy index. This probability is zero for the IBM, and it gets smaller values (asymptotically zero) for the PRM with higher $d_t$, see Fig.~\ref{subfig: uWave_Omni_Rayleigh__r_PfaPmd}.
%----------------------------figure-------------------------------
%-----------------------------------------------------------------

\begin{figure}[!t]
  \centering
  \includegraphics[width=\columnwidth]{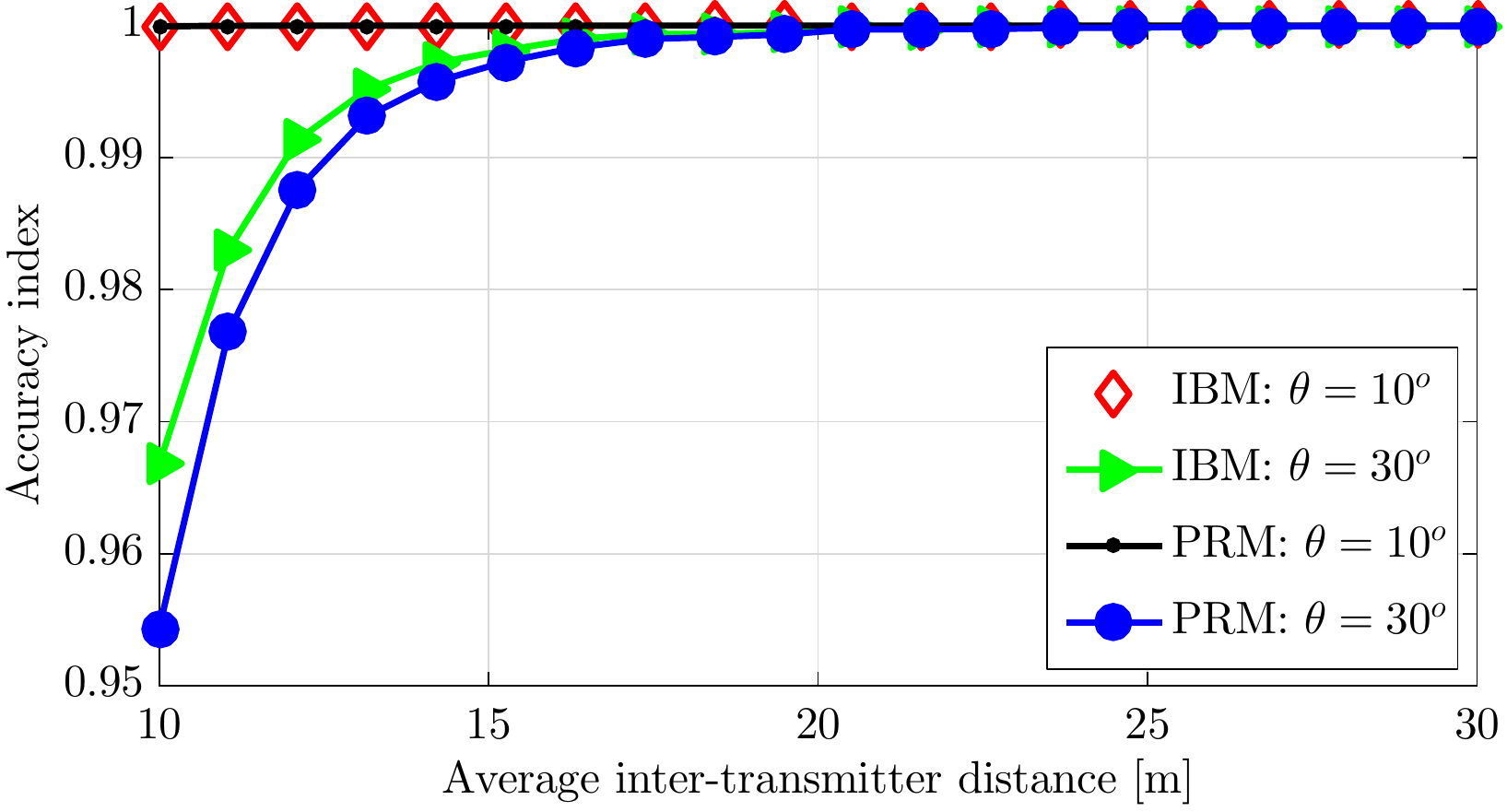}

  \caption{Accuracy of IBM and PRM under deterministic channel and directional communications. $r_{\text{PRM}} = \zeta^{-1/\alpha}$ where $\zeta$ is given in~\eqref{eq: zeta}, and $r_{\text{IBM}} = 2 r_{\text{PRM}}$.}
  \label{mmWave_Direc_Deterministic__lambdaT}
\vspace{-2mm}
\end{figure}

Fig.~\ref{mmWave_Direc_Deterministic__lambdaT} illustrates the impact of the operating bandwidth and average inter-transmitter distance on the accuracy index of both IBM and PRM under Scenario~2. As expected, directionality and blockage improve the accuracy of both interference models. Surprisingly, \emph{the PRM is accurate enough to motivate adopting this model to analyze and design of mmWave networks instead of the PhyM or even the IBM.}
%For relatively pencil-beams (e.g., $\theta=10 \degree$), which may be used in wireless backhauling applications, the accuracy of PRM in detecting outage events is almost 1 in all our simulations.
Compared to the PRM, the PhyM and IBM respectively have less than 5\% and 2\% higher accuracy in modeling the interference and detecting the outage events, but with substantially higher complexities. These complexities often result in limited (mostly intractable) mathematical analysis and little insight.
This highlights the importance of having quantitative (not only qualitative) insight of the accuracy of different interference models we may face in different wireless networks. Thereby, we can adopt a simple yet accurate enough model for link-level and system-level performance analysis.
%----------------------------figure-------------------------------

\section{Conclusions}\label{sec: conclusion}
We addressed very fundamental questions in analysis and design of wireless networks: how accurate different interference models are and how to select the right one. In particular, we proposed a new index that assesses the accuracy of any interference model in detecting outage events, under any set of assumptions on the communication protocols. Based on this index, we evaluated the accuracy of two prominent interference models, namely the classical protocol and interference ball models. Our detailed analysis revealed that, unlike the protocol model, the interference ball model can be arbitrary accurate by adding complexity into the model. Moreover, blockage and directionality can substantially improve the accuracy of both interference models. In such settings, even the simplest interference model may be almost as accurate as the most complex one. This is a promising feature of many future wireless technologies such as millimeter wave networks, which exhibit such blockage and directionality requirements, to significantly improve the mathematical tractability with negligible drop in the interference model accuracy.
\bibliographystyle{IEEEtran}
\bibliography{References}

\end{document}